# Topological stability versus thermal agitation in a metastable magnetic skyrmion lattice

H. Oike[1], A. Kikkawa[1], N. Kanazawa[2], Y. Taguchi[1], M. Kawasaki[1,2], Y. Tokura[1,2], and F. Kagawa[1,3†]

[1] *RIKEN Center for Emergent Matter Science (CEMS), Wako 351-0198, Japan*

[2] *Department of Applied Physics, The University of Tokyo, Tokyo 113-8656, Japan*

[3] *CREST, Japan Science and Technology Agency (JST), Tokyo 102-0076, Japan*

† To whom correspondence should be addressed. E-mail: fumitaka.kagawa@riken.jp

**Topologically stable matters can have a long lifetime, even if thermodynamically costly, when the thermal agitation is sufficiently low[1,2]. A magnetic skyrmion lattice (SkL) represents a unique form of long-range magnetic order that is topologically stable[3-9], and therefore, a long-lived, metastable SkL can form. Experimental observations of the SkL in bulk crystals, however, have mostly been limited to a finite and narrow temperature region in which the SkL is thermodynamically stable[5,7,10-14]; thus, the benefits of the topological stability remain unclear. Here, we report a metastable SkL created by quenching a thermodynamically stable SkL. Hall-resistivity measurements of MnSi reveal that, although the metastable SkL is short-lived at high temperatures, the lifetime becomes prolonged (>> 1 week) at low temperatures. The manipulation of a delicate balance between thermal agitation and the topological stability enables a deterministic creation/annihilation of the metastable SkL by exploiting electric heating and subsequent rapid cooling, thus establishing a facile method to control the formation of a SkL.**

The perspective of topology provides useful insights into the lifetime of a many-body system that is not in the most thermodynamically stable form. In an aggregate of carbon atoms, for instance, whereas the most thermodynamically stable form at standard temperatures and pressures is graphite, other forms, such as diamond and fullerene-molecule crystals, also have long lifetimes, despite a thermodynamic free energy penalty[15]. From a topological perspective[16], the carbon networks of the crystalline allotropes are topologically distinct from each other. Therefore, diamond and fullerene-molecule crystals are unable to transform into graphite unless thermal agitation enables the breaking and reformation of the carbon networks. States with this absence of continuity in topology are often termed topologically protected, or topologically stable, and this condition guarantees a finite lifetime for the thermodynamically less stable states of matter, thereby qualitatively accounting for their metastable nature.

Despite the many examples of topologically distinct crystals[16], only recently has a topologically stable long-range magnetic order been identified experimentally in the form of a magnetic SkL, which is a hexagonal arrangement of magnetic vortex lines[5-9] (for the schematic spin configuration, see Fig. 1a). Because of its topology, the SkL is topologically protected from other competing non-topological spin configurations, such as ferromagnetic and spin-conical orders (Fig. 1a), thus providing a unique platform to observe how long a topologically stable long-range ordering of spins can subsist beyond its thermodynamic stability. To address this issue, the lifetime of a metastable SkL requires investigation. In reality, however, unless randomly positioned dopant atoms[6] or pressure inhomogeneities[17] are introduced, the SkL in various chiral magnets is readily annihilated when the SkL becomes thermodynamically costly by passing through the SkL-to-conical transition line[5,7,10-14] (for a typical phase diagram, see Fig. 2a). Thus, the intrinsic lifetime of the metastable SkL remains unknown.

In pursuing a metastable SkL that should potentially form even in a disorder-free system, we noted that in the creation process of natural diamonds, thermodynamically stable diamonds grown in the mantle are quenched when they move to the surface of the earth, and hence, the diamond-to-graphite transition is kinetically avoided. Based on this realization, we envisaged that the SkL-to-conical transition, which normally occurs during cooling, can also be kinetically avoided if the cooling rate is sufficiently high (Fig. 1a). To test this working hypothesis, we chose nominally dopant-free MnSi as an archetypal system of a skyrmion-hosting material and exploited rapid cooling following electric-pulse heating applied to the sample. For instance, when a rectangular current pulse of $3.1\times10^6$ Am$^{-2}$ with a duration of 100 ms is applied to the sample at 10 K, the sample temperature is observed to quickly increase and reach a quasi-steady state with a temperature of ≈33–34 K. After the pulse ceases, the sample is rapidly cooled to the environmental temperature at a cooling rate of ≈700 Ks$^{-1}$ (for the estimation of the time-varying sample temperature and the cooling rate, see Supplementary Information and Supplementary Fig. 1).

Using the quenching technique described, the impact of quenching on the magnetic structure can be explored beyond the normally used cooling rate, $10^{-2}$-$10^{0}$ Ks$^{-1}$. To this end, we focused on the Hall resistivity, $\rho_{yx}$, as a sensitive probe for SkL formation, particularly in MnSi[18,19], and found that $-\rho_{yx}$ at 10 K under 0.22 T indeed changes from –2 to 32 nΩcm after applying the current pulse described. Whereas the enhanced $\rho_{yx}$ value remains constant as a function of time (verified up to one week), it exhibits transition-like behaviour in response to a slow magnetic field sweep (5.0×10$^{-4}$ Ts$^{-1}$) and then decreases to the values corresponding to the equilibrium state (Fig. 1b). These observations are reasonably understood by considering that a long-lived metastable magnetic structure is formed as a result of heating and subsequent rapid cooling.

To elucidate the origin of the metastable state, we first examined whether it is actually a quenched state. We tailored the cooling rate with ramp pulses (for details, see Supplementary Information and Supplementary Fig. 1) and observed how the $\rho_{yx}$ value changes. Figure 1c summarizes the $\rho_{yx}$ values measured after applying various cooling rates; remarkably, sharp crossover behaviour is observed from the slowly cooled to the rapidly cooled regimes, thus verifying that the enhanced $\rho_{yx}$ value of ≈32 nΩcm is indeed a consequence of the quenching. Because neither the ordinary Hall nor the anomalous Hall resistivities can explain such a greatly enhanced value of $\rho_{yx}$ (Fig. 1b), it is important to consider the contributions of the topological Hall resistivity, which is known to arise in the presence of the SkL because of its real-space Berry phase associated with the nonzero topological winding number[18-21]. In fact, the magnitude of the enhancement, $\Delta\rho_{yx} \approx 31-34$ nΩcm, is similar to the reported value, 35–37 nΩcm, observed when the metastable SkL subsists at 10 K in MnSi under the influence of pressure inhomogeneity[17], which indicates that the SkL is the most likely quenched state.

This scenario is further supported by the observation of an enhanced $\rho_{yx}$ value under various magnetic field/current density conditions. The contour plot of $\rho_{yx}$ measured after the application of a pulse with a given current density is displayed in Fig. 1d; here, we find that the enhancement of $\rho_{yx}$ is limited to a magnetic field range of 0.17–0.26 T, which is in good agreement with that of the equilibrium SkL phase in our sample, 0.14–0.27 T (Fig. 2a). Furthermore, the enhancement of $\rho_{yx}$ becomes appreciable only if the deduced sample temperature during the pulse application (see the upper horizontal axis of Fig. 1d) reaches or exceeds the temperatures of the equilibrium SkL (27.0–28.8 K). These findings lead us to conclude that the quenched magnetic state is a metastable SkL caused by the kinetic avoidance of the SkL-to-conical transition. Unexpectedly to us, the metastable SkL is formed even when the sample is quenched from the paramagnetic regime (>28.8 K), indicating that the paramagnetic-to-SkL transition has faster kinetics than the SkL-to-conical transition and therefore cannot kinetically be avoided, at least at the present cooling rate, 700 Ks$^{-1}$.

Having established a method to create a metastable SkL in a nominally clean sample, we can now address the issue of the stability beyond the equilibrium phase diagram. We found that the long-lived metastable SkL can be created at any temperature between 3 and 23 K using the same quenching method (see Supplementary Fig. 2); accordingly, we could further examine the stability of the metastable SkL against a magnetic field sweep at each temperature and construct a magnetic phase diagram that includes the quenched SkL, as shown in Fig. 2b. Here, two important aspects can be highlighted. First, the temperature/magnetic-field region in which the quenched SkL can subsist is significantly extended compared to that of the equilibrium SkL phase, which exemplifies the remarkable stability of the metastable SkL. Second, there is a temperature gap (~24–27 K) in which neither the equilibrium SkL nor the metastable SkL can subsist, implying that the metastable SkL is short-lived in this temperature range.

Such a short-lived metastable SkL can be substantiated by the time decay of the enhanced $\rho_{yx}$ value, as shown in Fig. 3a. The relaxation of $\rho_{yx}$ to the equilibrium value can clearly be seen, in accord with the expectation that the metastable SkL is short-lived in this temperature range. The behaviour is characterized well by the standard relaxation equation (denoted by the broken lines in Fig. 3a):

$$\rho_{yx}(t) = \rho_{yx,0} + (\rho_{yx,\infty} - \rho_{yx,0})\{1 - \exp(-t/\tau)\}, \qquad (1)$$

where $\rho_{yx,0}$ and $\rho_{yx,\infty}$ denote the initial and fully relaxed (or equilibrium) values of $\rho_{yx}$, respectively, and $\tau$ represents the relaxation time (or, equivalently, the lifetime of the metastable SkL). Figure 3b summarizes the $\tau$ values derived for various temperatures. Note that as the temperature approaches the SkL transition temperature of 27 K, the relaxation time decreases dramatically and will likely decrease to less than 10 s. This behaviour reasonably accounts for why the SkL is readily annihilated during cooling in nominally clean systems, even though it is topologically stable[5,7,10-14]. Nevertheless, after quickly passing through the high-temperature region in which the metastable SkL is short-lived, the lifetime becomes unmeasurably long toward low temperatures. Consequently, the quenched SkL is stabilized practically. The significant slowing down of $\tau$ invokes thermal activation process and suggests that the SkL-to-conical topology change proceeds via an intermediate spin configuration with a creation energy, $E_g$. In estimating the order of magnitude of $E_g$, we tentatively assume that $E_g$ is temperature-dependent and proportional to the square of the local magnetic moment, $m$. Then, if we adopt a simplified $m$-temperature profile $m \sim (T_c-T)^{0.5}$ with $T_c \approx 29$ K (the SkL-paramagnetic transition temperature), $E_g$ at zero temperature is estimated to be $\sim 2 \times 10^3$ K (for details, see Supplementary Fig. 4). Although the intermediate spin configurations cannot be described clearly based on the $\rho_{yx}$ data, the "emergent magnetic monopoles" discussed in the literature[22] is likely.

Finally, based on the present findings, we demonstrate reversible and deterministic switching between the topologically distinct spin configurations. Figure 4a shows phenomenological free-energy landscapes of the relevant spin configurations, illustrating the working principles of the switching. In the "SET process", as has been discussed in Figs. 1b-d, the application of a current pulse to the conical state results in the metastable SkL via the kinetic avoidance of the SkL-to-conical transition. In the "RESET" process, a relatively long current pulse (10 s) with a moderate intensity ($1.7 \times 10^6$ Am$^{-2}$) was applied to the quenched SkL to heat the sample to 25–27 K, at which temperature the metastable SkL is short-lived (<10 s; see Fig. 3b) and can therefore relax into the conical state within the pulse duration. The reversible electric switching of the topological/non-topological magnetic structure is thus feasible, as manifested in the switching of $\rho_{yx}$ (Figs. 4b,c). The switching exhibits good reparability (Fig. 4d), highlighting the deterministic nature of the working principle; that is, whereas thermal agitation at high temperatures allows the system to relax into the most thermodynamically stable state within a relatively short time, such relaxation is almost unexpected at low temperatures because the topological stability surpasses the thermal agitation.

Our observations clearly establish that a metastable SkL hidden behind non-topological magnetic ordering can be accessed by applying sufficiently high cooling rates. Because the local moments generally rise as the temperature decreases, SkL-derived physical properties become more accessible in the quenched SkL, facilitating experimental studies. In the context of the possible application of skyrmions, in which the manipulation of a single skyrmion is ultimately considered[23-27], we conjecture that a similar creation/annihilation method is additionally applicable to the case of a small number of skyrmions because the working principle demonstrated above is predicated on manipulating the delicate balance among the topological stability, the thermodynamic stability, and the thermal agitation, properties that remain well defined in a finite-size system. We have also shown that the topological stability is undermined at high temperatures, suggesting that the operating temperature of skyrmion devices must be chosen in the context of a competition between the topological stability and the thermal agitation.

**Methods**

**Sample preparation**

A single crystal of MnSi was grown by the Czochralski method. The sample was oriented by Laue X-ray diffraction, cut with a wire saw, and polished to a size of $2.5 \times 0.95 \times 0.1$ mm$^3$, with the largest surface normal to the <100> axis. The residual resistivity ratio of the sample used in this study was ≈55. Cu current leads of 0.43 mm$\phi$ were attached to the sample and fixed with silver paste. Au wires for the voltage probe were soldered to the sample with indium.

**Transport measurements**

The Hall resistivity $\rho_{yx}$ was measured at 33 Hz of the current excitation with the five-probe method under a magnetic field parallel to the <100> axis using a lock-in amplifier (Signal Recovery, 7270 DSP) equipped with a transformer preamplifier (Stanford Research Systems, SR554). The current density of the a.c. excitation was set to less than $1.4 \times 10^5$ Am$^{-2}$. We confirmed that the a.c. current caused no appreciable increase in the sample temperature. The pulse currents used for the sample heating were generated by a function generator (NF Corporation, WF1947) connected with a bipolar amplifier (NF Corporation, HSA4014). The voltage drop between the voltage probes attached to the sample was amplified with a low-noise preamplifier (NF Corporation, LI-75A) and was monitored as a function of time using a data logger (Data Translation, DT8824), so that we could derive the time-varying sample temperature (see Supplementary Information). During the time-varying Hall resistivity measurements, we monitored the digital-to-analog convertor output of the lock-in amplifier using the data logger.

**Acknowledgements** We thank N. Nagaosa, W. Koshibae and A. Rosch for fruitful discussions. This work was partially supported by JSPS KAKENHI (Grant Nos. 25220709, 24224009,15H05459).

**Author Contributions** H.O. conducted all experiments and analyzed the data. A.K. grew the single crystals used for the study. F.K. planned and supervised the project. H.O. and F.K. wrote the letter. All authors discussed the results and commented on the manuscript.

**Additional Information** Supplementary Information is available in the online version of the paper. Reprints and permissions information is available online at www.nature.com/reprints. Correspondence and requests for materials should be addressed to F.K.

**Competing financial interests** The authors declare no competing financial interests.

**Figure 1 │Metastable skyrmion lattice stabilized under rapid cooling. a,** The schematic of the cooling-rate-dependent bifurcation of an equilibrium skyrmion lattice (SkL). Whereas the equilibrium SkL changes into the thermodynamically stable conical phase under slow cooling, rapid cooling causes kinetic avoidance of the SkL-to-conical transition, resulting in a quenched metastable SkL. **b,** The magnetic field dependence of the Hall resistivity, $\rho_{yx}$, measured before and after quenching at 0.22 T. By applying a current pulse to the equilibrium state under 0.22 T (the cross), the $\rho_{yx}$ value changes to an enhanced value (the open circle). **c,** The cooling-rate dependence of $\rho_{yx}$ measured at 10 K and 0.22 T after quenching is complete. **d,** The contour plot of $\rho_{yx}$ measured at 10 K post-quenching using various current densities under selected magnetic fields. The deduced sample temperature during the electric current application is also indicated on the upper horizontal axis (see also Supplementary Fig. 1).

**Figure 2 │Magnetic phase diagrams of MnSi under the equilibrium and quenched conditions. a,** The thermodynamic equilibrium phase diagram. **b,** The phase diagram including the metastable SkL created by quenching under a magnetic field. Red symbols in **b** were determined from the data shown in Fig. 1b and

Supplementary Fig. 2. Whereas the boundaries of the equilibrium SkL were determined from the $\rho_{xx}$ data, the ferromagnetic transition line was determined from the $\rho_{yx}$ data (see Supplementary Figs. 2 and 3).

**Figure 3 │ Lifetime of the metastable skyrmion lattice. a,** The time evolution of $\rho_{yx}$ at selected temperatures after quenching at 0.22 T. The broken red lines are fits to the standard relaxation equation; see Eq. (1) in the text. **b,** The temperature dependence of the lifetime of the metastable SkL at 0.22 T. The lifetimes are obtained from the fitting results in **a**.

**Figure 4 │ Creation and annihilation of the metastable skyrmion lattice through thermal control. a**, A scheme for the repeatable switching between the SkL and the conical phase in terms of a schematic free-energy landscape in a multidimensional spin-configuration space. Note that, whereas the SkL is thermodynamically stable in the red-hatched temperature region (27–29 K), the conical state is thermodynamically stable in the blue-hatched temperature region (<27 K). **b, c,** Single-cycle operation of the switching under the application of rectangular current pulses. The time profiles of the $\rho_{yx}$ value and current are shown in **b** and **c**, respectively. **d,** Repetitive switching between the metastable SkL and the equilibrium conical phases.

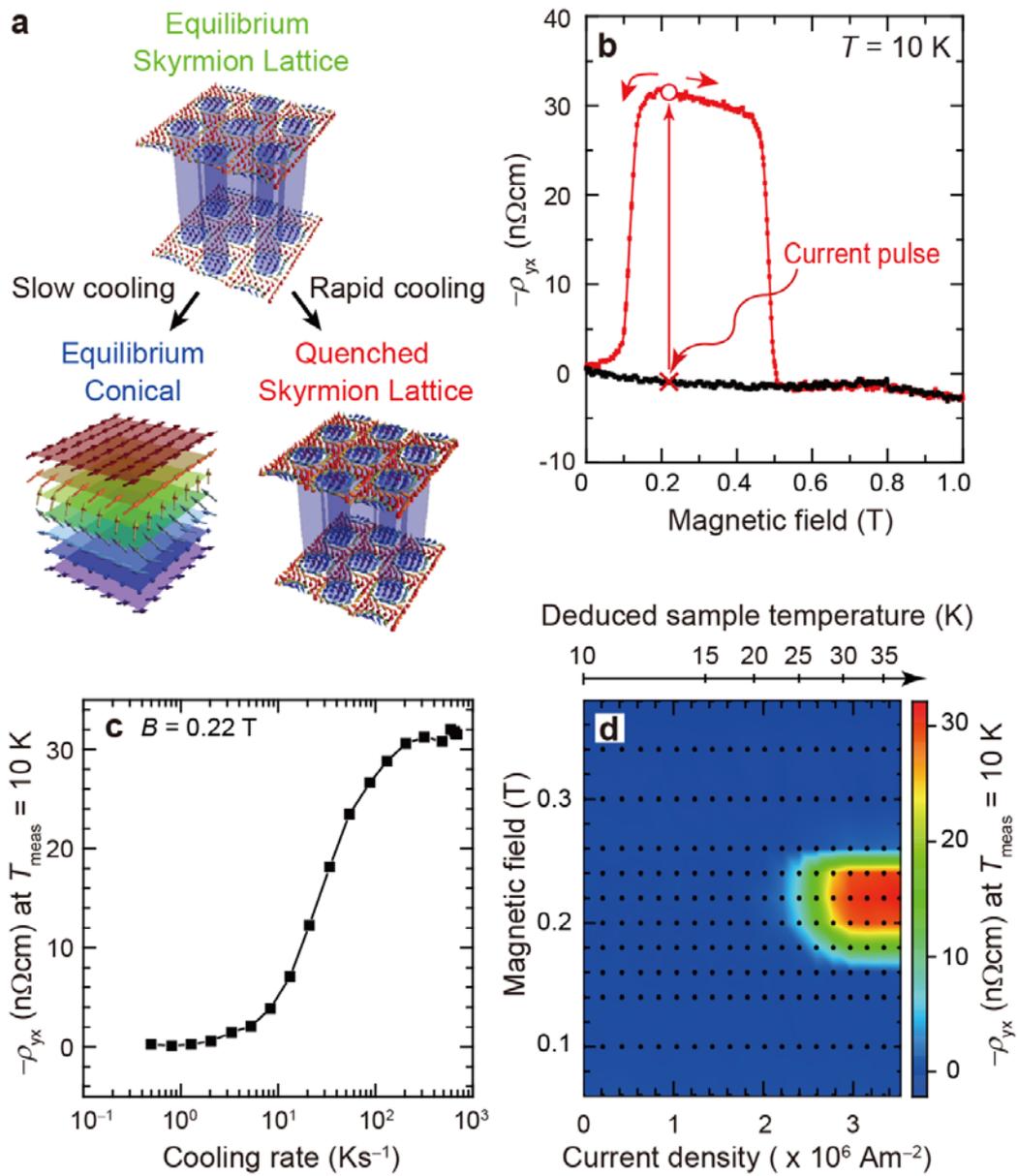

Fig. 1

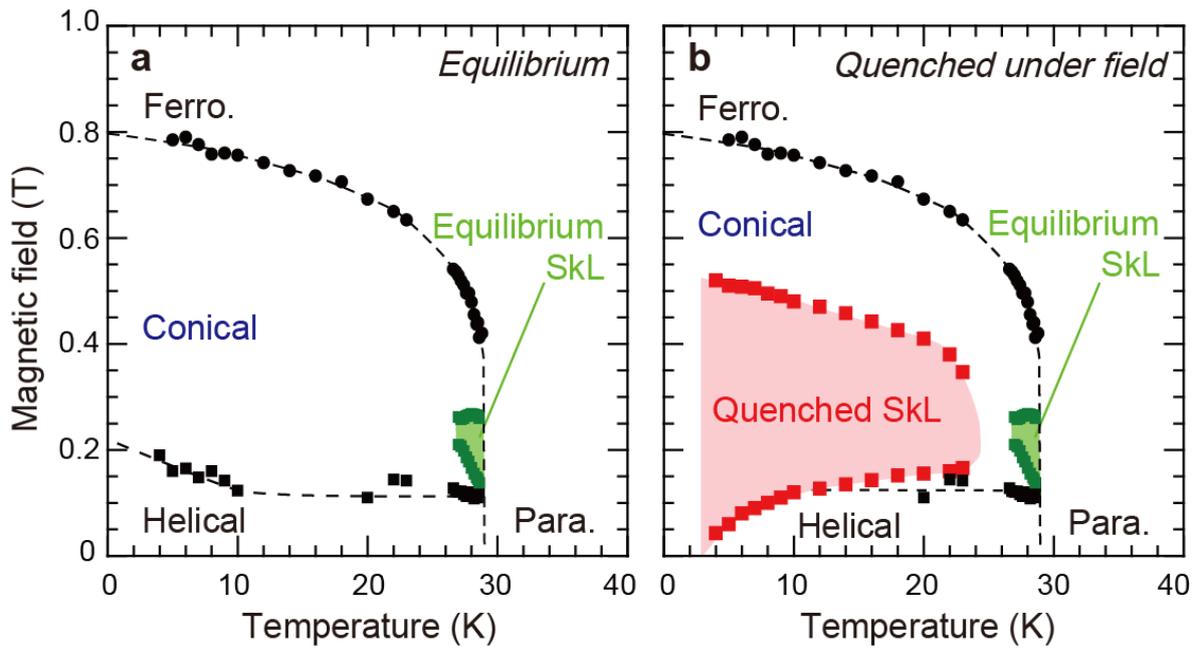

Fig. 2

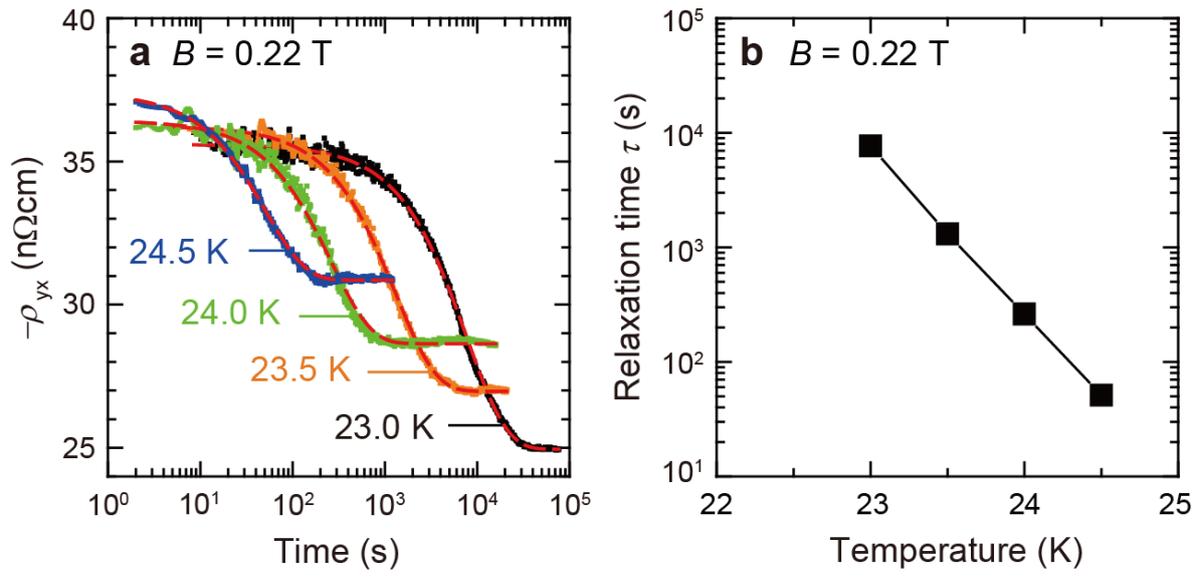

Fig. 3

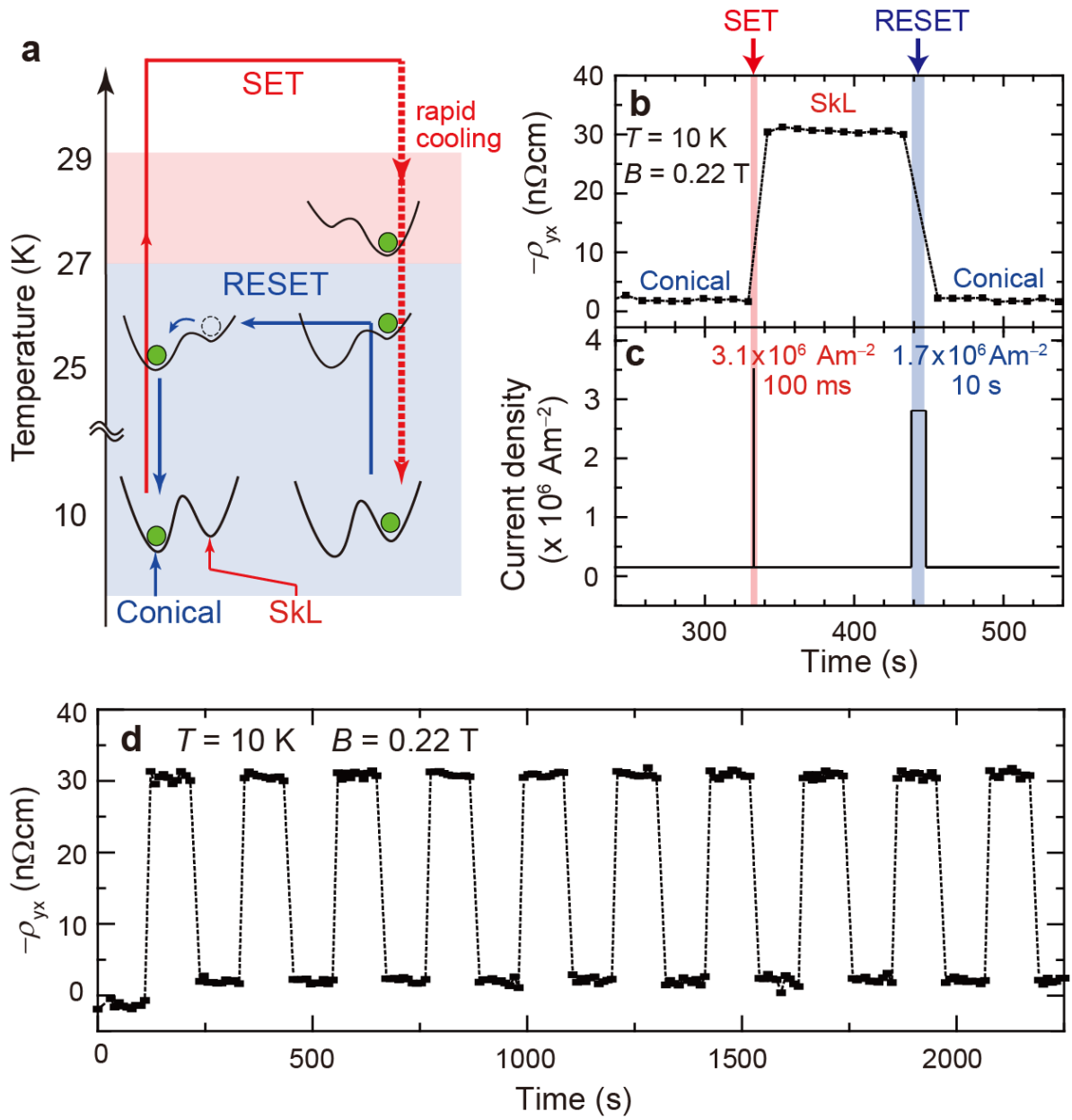

Fig. 4

# Supplementary Information for

# Topological stability versus thermal agitation in a metastable magnetic skyrmion lattice

H. Oike[1], A. Kikkawa[1], N. Kanazawa[2], Y. Taguchi[1], M. Kawasaki[1,2], Y. Tokura[1,2], and F. Kagawa[1,3†]

[1] *RIKEN Center for Emergent Matter Science (CEMS), Wako 351-0198, Japan*

[2] *Department of Applied Physics, The University of Tokyo, Tokyo 113-8656, Japan*

[3] *CREST, Japan Science and Technology Agency (JST), Tokyo 102-0076, Japan*

† To whom correspondence should be addressed. E-mail: fumitaka.kagawa@riken.jp

**Quenching method**

As described in the main text, we have achieved a cooling rate of up to 700 Ks$^{-1}$ by exploiting rapid cooling following pulsed Joule heating. In Supplementary Information, we detail how we controlled and estimated the cooling rates. Figures S1a-c display the typical pulse shapes used in the study: we chose 100 ms as the minimum pulse duration so that the sample temperature approximately reached a steady state under the current application (for instance, see Fig. S1g); then, the current was decreased at a given ramp rate, facilitating control of the cooling rate. To estimate the cooling rate, the time evolution of the sample temperature was deduced from that of the longitudinal resistivity (Figs. S1d-f), $\rho_{xx}$, by referring to the $\rho_{xx}$-temperature profile (Fig. S1j). Figures S1d-f and S1g-i display the time evolution of $\rho_{xx}$ and the deduced sample temperatures, respectively, for the case of the environmental temperature of 10 K. For instance, when a current pulse with the maximum current density $J = 3.1\times10^6$ Am$^{-2}$ and a ramp rate of $-3.1\times10^8$ Am$^{-2}$s$^{-1}$ were applied, a cooling rate of 700 Ks$^{-1}$ was achieved upon passing through the SkL-to-conical transition temperature, $\approx 27$ K (Fig. S1g). When the ramp rate was decreased to $-3.1\times10^7$ and $-6.2\times10^6$ Am$^{-2}$s$^{-1}$ (Figs. S1b,c), the deduced cooling rates accordingly decreased to 300 and 70 Ks$^{-1}$, respectively (Figs. S1h,i). In this way, manipulation of the ramp rate allows us to achieve any cooling rate less than 700 Ks$^{-1}$, as shown in Fig. S1k. Note that, although the cooling rate increases approximately linearly with the ramp rate in the slow ramp-rate regime, it saturates at $\approx 700$ Ks$^{-1}$ in the high ramp-rate regime because of the finite speed of the thermal diffusion.

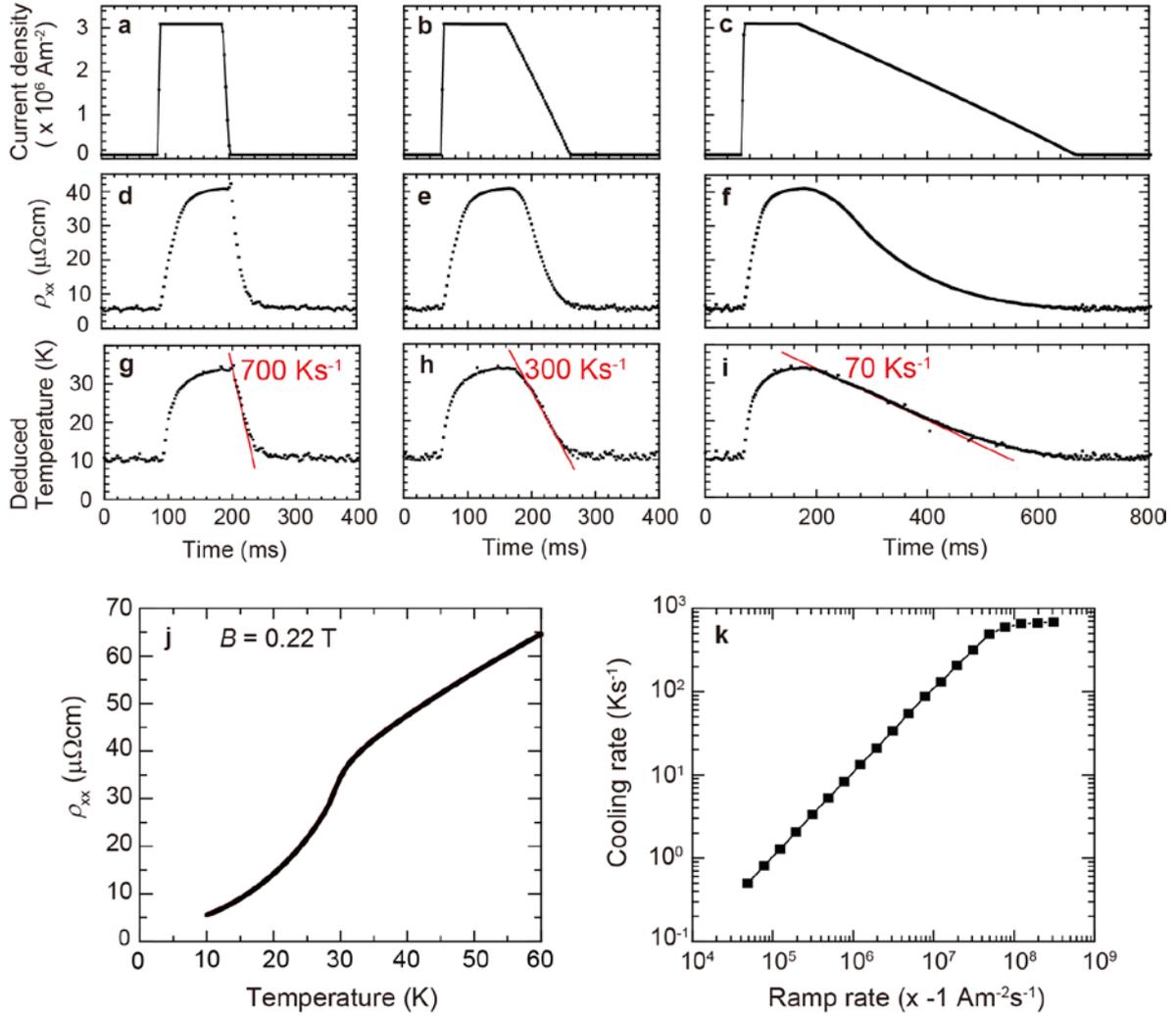

**Figure S1 | Methods to control the cooling rate following Joule heating**. **a-i,** The time profiles of the current density (**a-c**), the longitudinal resistivity $\rho_{xx}$ (**d-f**), and the deduced sample temperature (**g-i**). The ramp rate of the current-decrease process is $-3.1\times10^8$ (for **a, d, g**), $-3.1\times10^7$ (for **b, e, h**), and $-6.2\times10^6$ Am$^{-2}$s$^{-1}$ (for **c, f, i**). The red lines represent the cooling rate at approximately the SkL-to-conical transition temperature, ≈27 K. **j,** The temperature dependence of $\rho_{xx}$ under 0.22 T, from which we deduced the time profiles of the sample temperature (**g-i**). **k,** The achieved cooling rate at approximately 27 K versus the ramp rate used for the current-decrease process. All the data in **a-k** were obtained at 0.22 T and an environmental temperature of 10 K.

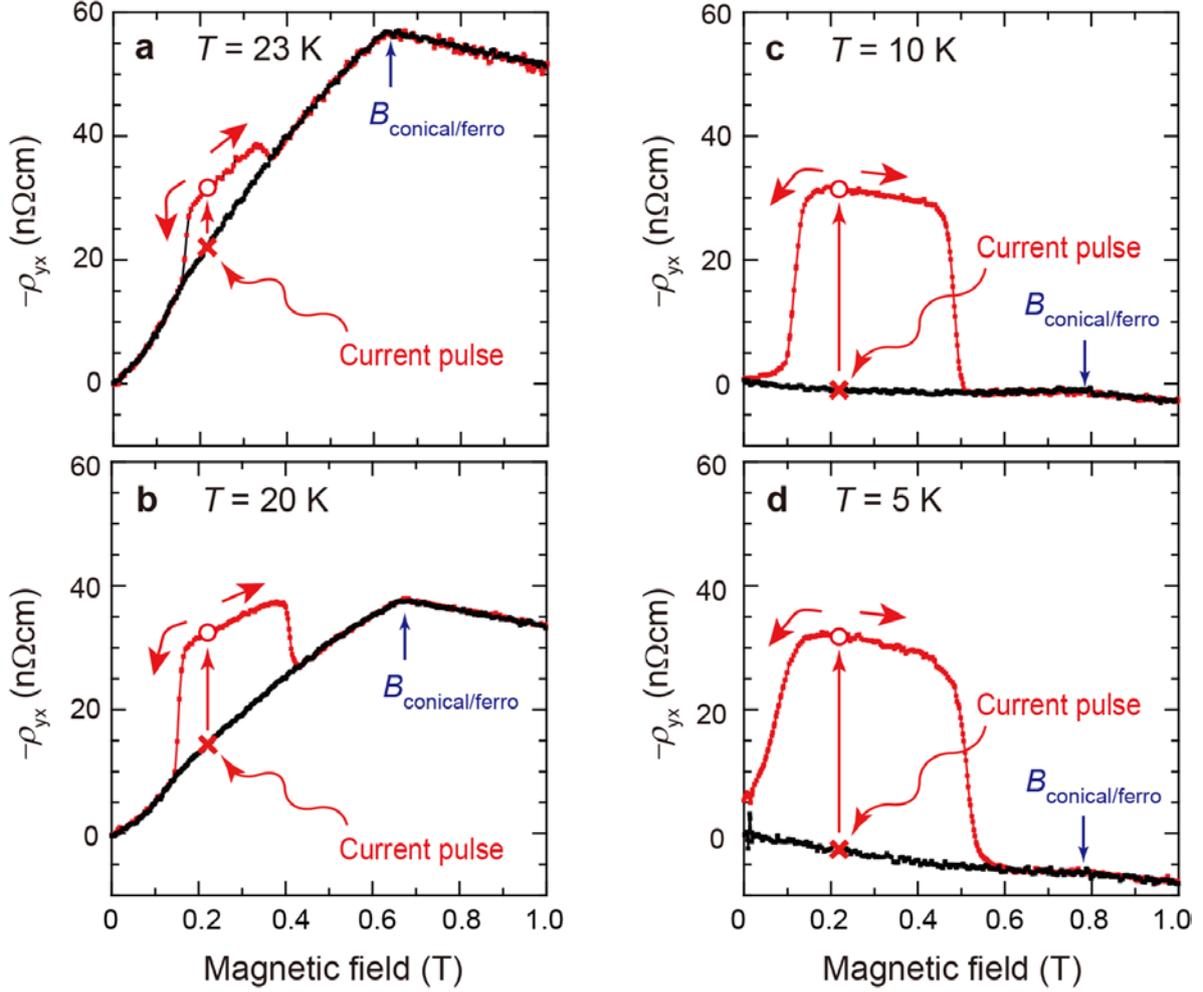

**Figure S2 | Stability of the metastable SkL against a magnetic field sweep**. **a-d,** The magnetic field dependence of $\rho_{yx}$ at 23 (**a**), 20 (**b**), 10 (**c**), and 5 K (**d**). The magnetic field sweep rate was fixed at $5.0\times 10^{-4}$ Ts$^{-1}$. By applying a current pulse to the equilibrium state at 0.22 T (the red cross), $\rho_{yx}$ exhibits an enhanced value (the red open circle). Then, when the magnetic field is either increased or decreased from 0.22 T, $\rho_{yx}$ sharply decreases and coincides with the values corresponding to the equilibrium state. We regarded the midpoint of the sharp drop of $\rho_{yx}$ as the critical magnetic field at which the metastable SkL converts into the equilibrium conical/helical phase.

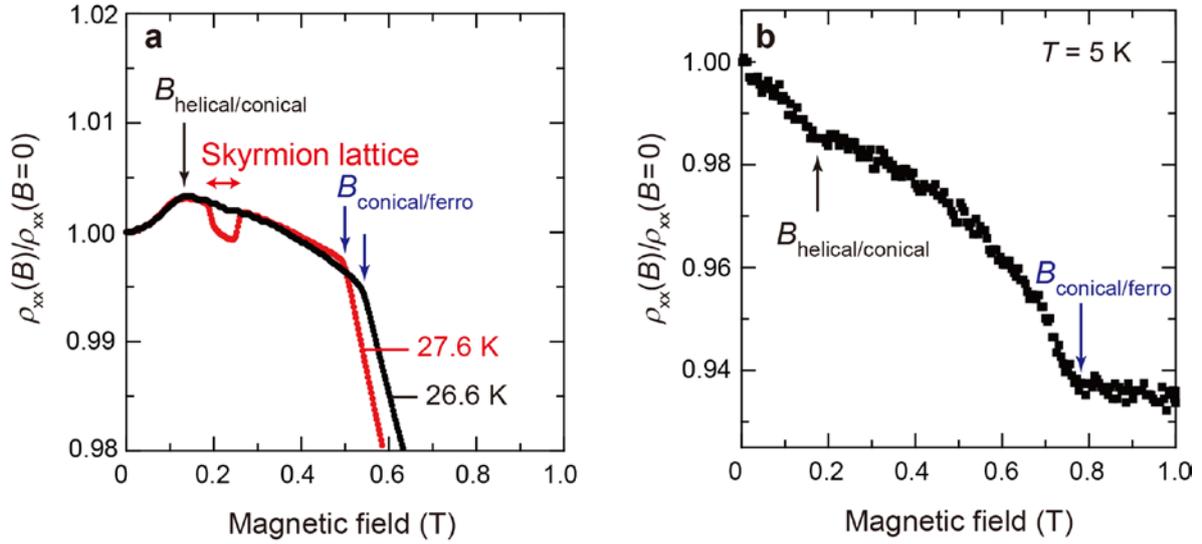

**Figure S3 | Magnetotransport data at selected temperatures. a,b,** The magnetic field dependence of the normalized $\rho_{xx}$ value at 27.6 and 26.6 K (**a**) and at 5 K (**b**). As the magnetic field increases, the slope of the magnetoresistance changes successively upon passing through the transition magnetic fields from the helical phase to the conical phase ($B_{helical/conical}$) and from the conical phase to the ferromagnetic phase ($B_{conical/ferro}$). At 27.6 K, a dip in the $\rho_{xx}$ profile is also observed, indicating the magnetic field range of the equilibrium SkL phase.

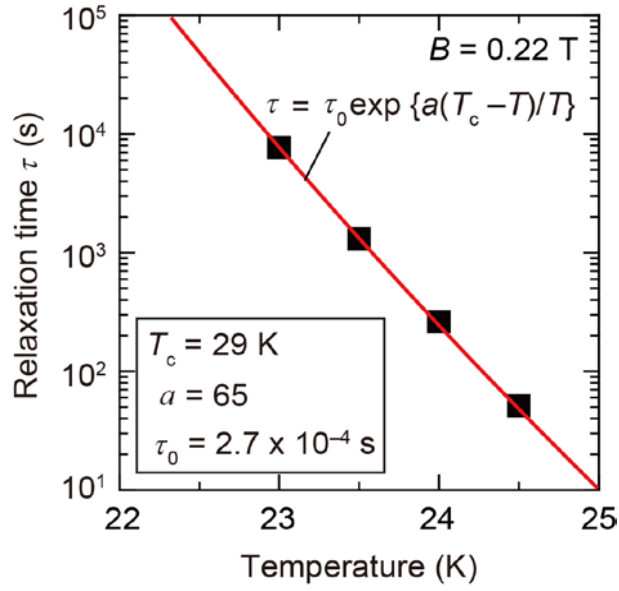

**Figure S4 | Analysis of the temperature dependence of $\tau$.** Several assumptions are required to estimate the creation energy $E_g$ of the intermediate spin configuration (see the text). For the order-of-magnitude estimate, we tentatively assumed that $E_g$ is temperature-dependent and proportional to the square of the local magnetic moment, $m$; furthermore, we adopted a simplified $m$–temperature profile $m \sim (T_c-T)^{0.5}$ with $T_c \approx 29$ K, the SkL-paramagnetic transition temperature. Thus, we fitted the temperature dependence of $\tau$ by using $\tau = \tau_0\exp(E_g/k_BT)$ with $E_g/k_B = a(T_c-T)$, where $\tau_0$ and $a$ represent proportional constants, and $k_B$ is Boltzmann's constant. From the fitting results (the red curve), we obtained $a \approx 65$. Hence, within the present approximation, $E_g$ at zero temperature was estimated to be $\sim 2\times 10^3$ K, whereas $E_g$, for instance, at 24 K was $\sim 3\times 10^2$ K.